\documentclass[12pt]{iopart}
\usepackage{graphicx}
\usepackage{amssymb}
\usepackage{epstopdf}
\begin{document}

\title{Spangolite: an $s=1/2$ maple leaf lattice antiferromagnet?}
\author{T Fennell$^{1}$\footnote{Began this work at The London Centre for Nanotechnology, 17-19 Gordon Street, London WC1H 0AH, UK}, JO Piatek$^2$, RA Stephenson$^3$, GJ Nilsen$^2$ and HM R{\o}nnow$^2$}
\address{$^1$ Institut Laue Langevin,  BP 156, 6, rue Jules Horowitz, 38042, Grenoble Cedex 9, France}
\ead{fennell@ill.fr}

\address{$^2$ Laboratory for Quantum Magnetism, \'Ecole Polytechnique F\'ed\'erale de Lausanne (EPFL), 1015 Lausanne, Switzerland}                                           
\address{$^3$ School of Chemistry, University of Southampton, Highfield, Southampton, SO17 1BJ, UK}

\begin{abstract}
Spangolite, Cu$_6$Al(SO$_4$)(OH)$_{12}$Cl$\cdot$3H$_2$O, is a hydrated layered copper sulphate mineral.  The Cu$^{2+}$ ions of each layer form a systematically depleted triangular lattice which approximates a maple leaf lattice.  We present details of the crystal structure, which suggest that in spangolite this lattice actually comprises two species of edge linked trimers with different exchange parameters.  However, magnetic susceptibility measurements show that despite the structural trimers, the magnetic properties are dominated by dimerization.  The high temperature magnetic moment is strongly reduced below that expected for the six $s=1/2$ in the unit cell.
\end{abstract}

\maketitle

Unconventional groundstates and excitations, combined with the possibility of direct connection with quantum many body theories, drive the study of low dimensional magnetic materials with $s = 1/2$ or $s = 1$ magnetic moments.  Geometrically frustrated magnets are also of interest as their macroscopic ground state degeneracy often results in unusual behaviour.  The combination of $s=1/2$ magnetic moments with a frustrated lattice is therefore particularly sought after.  Systems such as SrCu$_2$(BO$_3$)$_2$~\cite{Kageyama:1999p2064}, a good realization of the Shastry-Sutherland lattice, or the kagom{\'e} lattices found in materials such as Herbertsmithite~\cite{devries_2009} or Volborthite~\cite{Nilsen:2010p3168} presumably exemplify the tip of the iceberg in terms of model materials with $s=1/2$ magnetic moments on a frustrated lattice.     

Examples of frustrated lattice geometries now studied include the aforementioned Shastry-Sutherland lattice in SrCu$_2$(BO$_3$)$_2$~\cite{Kageyama:1999p2064}; triangular lattices of dimers in Ba$_3$Mn$_2$O$_8$~\cite{Stone:2008p1278}, Sr$_3$Cr$_2$O$_8$~\cite{singh,QuinteroCastro:2010p1983} and Cs$_3$Cr$_2$Br$_9$~\cite{Grenier:2006p474}; the triangular kagom{\'e} lattice realized in CuX(cpa)$_6$ (cpa = carboxypentonic acid, X$=$F, Cl, Br)~\cite{Maruti:1994p1237,Loh:2008p1361}; mutually perpendicular linear trimers in  2b$\cdot$3CuCl$_2\cdot$2H$_2$O (b = betaine, C$_5$H$_{11}$NO$_2$)~\cite{RemovicLanger:2009p2781}; or the triangular lattice of nonamers in La$_3$Cu$_2$VO$_9$~\cite{Robert:2008p1981}.  The properties of various spin Hamiltonians on common triangle based lattices (i.e. triangular or kagom{\'e}) are reviewed by Normand~\cite{normand_2009}, and those of the $s=1/2$ Heisenberg antiferromagnet on many lattices by Richter {\it et al.}~\cite{Richter:2004p3164}.  Included  in the latter survey are two maple leaf lattices, depleted triangular lattices constructed by decorating hexagons with edge-sharing triangles.  The $s=1/2$ Heisenberg antiferromagnet on the maple leaf lattice was found to have a six sublattice ordered ground state which persists in the presence of quantum fluctuations~\cite{Schmalfuss:2002p3338}.  No experimental realization with $s=1/2$ is known, although the series of M$_x$[Fe(O$_2$CCH$_2$)$_2$NCH$_2$PO$_3$]$_6$$\cdot$$n$H$_2$O (M = Na, $x$ = 
11; $M$ = K, $x$ = 11; $M$ = Rb, $x$ = 10) approximate maple leaf lattices with $s=5/2$~\cite{Cave:2006p6941}.

Amongst many layered copper minerals in the Inorganic Crystal Structure Database (ICSD)~\cite{icsd_ref}, we have identified spangolite~\cite{Anonymous:1893p3124,FRONDEL:1949p3126,Hawthorne:1993p1446} as a candidate for further investigation.  Spangolite, Cu$_6$Al(SO$_4$)(OH)$_{12}$Cl$\cdot$3H$_2$O, is a hydroxy-hydrated copper aluminum sulphate mineral  in which the copper ions form well separated, depleted triangular layers, approximating a maple leaf lattice.  Although it has been known to mineralogists for many years~\cite{Anonymous:1893p3124}, its crystal structure was only determined considerably more recently~\cite{FRONDEL:1949p3126,Hawthorne:1993p1446}, and its magnetic properties are unknown.

\section{Experimental}

We obtained specimens of spangolite originating from three different locations from two mineral suppliers: Blanchard mine, Socorro, Colorado (Dakota Matrix Minerals), Yerrington Mine, Nevada and Fontana Rossa, Corsica (both from Excaliber Mineral Corporation).  Spangolite is described as forming thin plates and has a turqouise colour.  The specimens generally carry more than one copper mineral and the material is often very fine grained.    After examining and measuring the susceptibility of various samples we found it best to confine our attention to the Blanchard Mine specimen which has distinguishable crystallites conforming to the description.  We extracted 19 mg from this specimen and examined every piece (of which there were about twenty) to verify that they all had the form and color described above.  The measured susceptibility was consistent with a single phase and a small part of this material was used to confirm the structure by single crystal x-ray diffraction.

The x-ray diffraction measurements were made using a Bruker-Nonius FR591 rotating anode diffractometer operating with Mo K$\alpha$ radiation ($\lambda = 0.711$ \AA) and equipped with a Bruker-Nonius Roper CCD camera and $\kappa$-goniostat driven by COLLECT~\cite{hooft}.  The sample was cooled to 120 K using a cryostream (Oxford Cryosystems Cobra).  26343 reflections were measured and combined to give 1308 independent  reflections.  The data was processed using the DENZO~\cite{denzo} software and corrected for absorption using SADABS~\cite{sadabs}, the $R$ factor for combination of reflections was $R=5.5$\%.  The crystal structure was determined using SHELXS-97 and refined using SHELXL-97~\cite{sheldrick}.  Non-hydrogen atoms were refined anisotropically and hydrogen atom positions and thermal parameters were fixed at idealized values riding on those of a parent atom.  A good fit was obtained with final values of $R_J = 4.45$\% and $R_W = 8.09$\%.

Susceptibility measurements were made using a Quantum Design MPMS SQUID magnetometer between 1.8 and 300 K in a field of 1000 G.  We measured the low temperature part of the susceptibility at temperatures between 0.05 and 3 K using an AC susceptometer (from Cambridge Magnetic Refrigeration) and a dilution
refrigerator (Oxford Instruments Kelvinox 25). The applied AC field had a frequency of 990 Hz and
amplitude of 42 mG. A constant background was subtracted from the data
and the resulting inverse susceptibility scaled and offset to overlap
with the high temperature data between 2 and 3 K.

\section{Results}

\begin{figure} 
\begin{center}
   \includegraphics[trim=0 0 0 0,clip=true,scale=0.64]{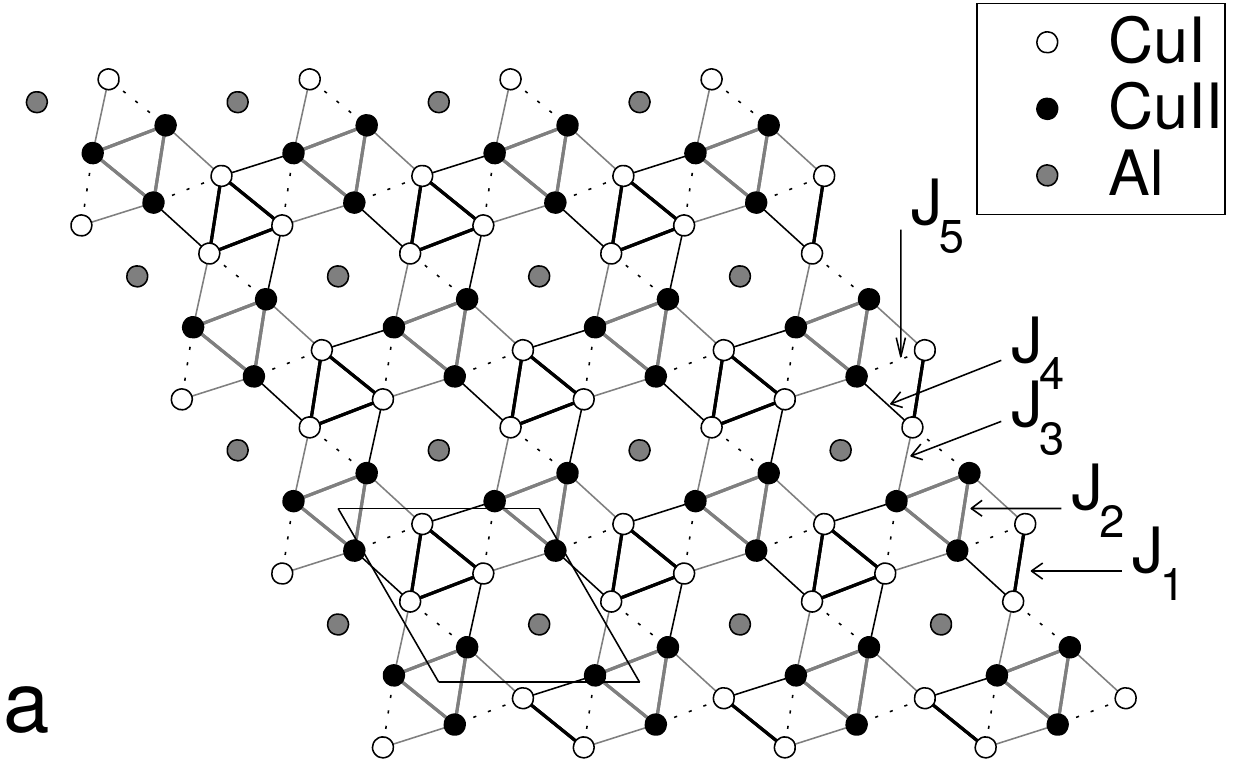} \\
   \vskip 0.4cm
   \includegraphics[trim=1 1 1 1,clip=true,scale=0.84]{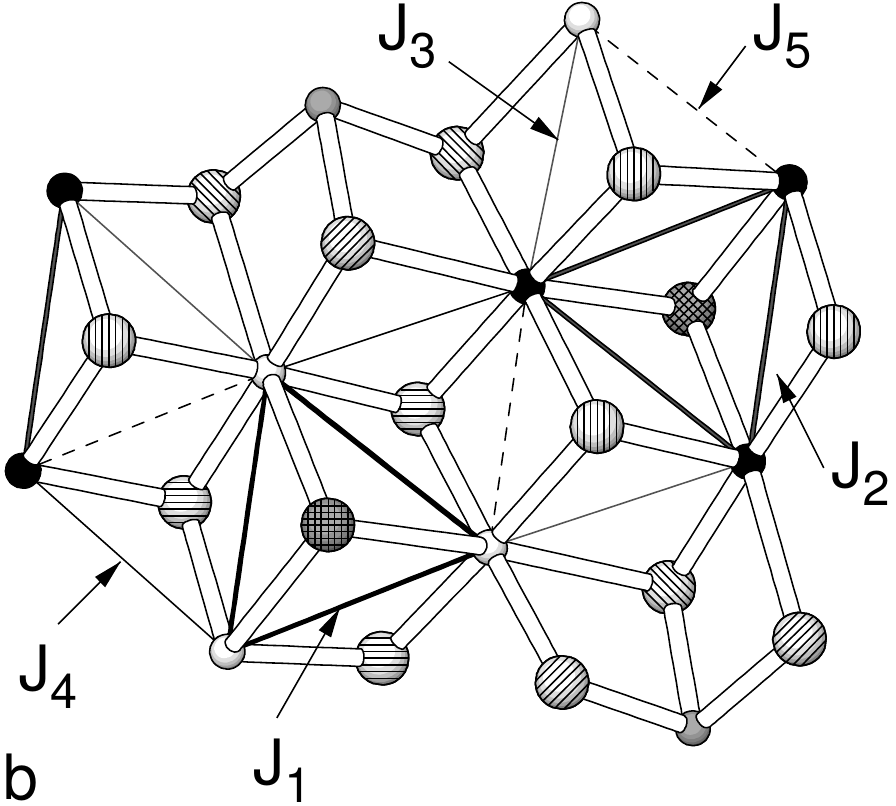}    
 \caption{a: The lattices of metal cations in spangolite.  The two copper sites CuI and CuII are shown, with their near neighbor contacts.  It can be seen that they form two families of pure trimers (bold lines) which are linked by two other types of trimer (fine lines).  According to the Goodenough-Kanamori rules, all the bonds are antiferromagnetic, except the dashed ones, which will be ferromagnetic.  The lattice is derived from a triangular lattice, with $1/7$ depletion introduced by the triangular superlattice of aluminum atoms and slight distortion to produce the different families of trimer. b: The crystallographic structure of spangolite.  The cations are shown as small spheres with the same colors as the top panel, the trimers are shown with the same scheme as the top panel.  The anions are shown as large spheres with the following codes: Cl$^-$, square hatching; O(SO$_3$)$^{2-}$, (i.e. O2) cross hatching (only the bridging oxygen atom is shown, the rest of the group projects below the plane); O3, diagonal downward hatching; O4, diagonal upward hatching; O5, horizontal hatching; O6, vertical hatching.  No hydrogen atoms or interlayer waters are shown. }
   \label{spanglatt}
   \end{center}
\end{figure}

\fulltable{ \label{table1}First neighbour superexchange paths, Cu-Cu distances and bridging angles in spangolite.}    
   \scriptsize{\begin{tabular}{@{}ccccccc} 
      \br
Name & Sites &  Path & Distance (\AA) & Angle (degrees) & G-K & expected total exchange\\
   & & & & & exchange & + color in Fig.~\ref{spanglatt}\\
\mr
$J_1$&CuI$-\mathrm{O}_5-$CuI      & Cu-O(H)-Cu    & 3.214  & 109.53$^\circ$ & AFM & AFM\\
 &CuI$-\mathrm{Cl}-$CuI      & Cu-Cl-Cu    &  & 67.68$^o$ & FM & black (bold)\\
      \mr
$J_2$&CuII$-\mathrm{O}_6-$CuII &      Cu-O(H)-Cu& 3.213 & 108.42$^\circ$ & AFM &  AFM\\
 &CuII$-\mathrm{O}_2-$CuII      & Cu-O(SO$_3$)-Cu   &   & 83.77$^\circ$ & FM & gray (bold)\\
      \mr
$J_3$&CuI$-\mathrm{O}_6-$ CuII    &  Cu-O(H)-Cu    &  3.107 & 106.50$^\circ$ & AFM  & AFM\\   
 &CuI$-\mathrm{O}_3-$CuII      & Cu-O(H)-Cu    &  & 89.69$^\circ$ & FM & gray (fine)\\
\mr
$J_4$&CuI$-\mathrm{O}_5-$CuII     & Cu-O(H)-Cu    & 3.111 & 103.37$^\circ$ & AFM & AFM\\
  &CuI$-\mathrm{O}_4-$CuII     &  Cu-O(H)-Cu    &   & 93.23$^\circ$ & FM & black (fine)\\
      \mr
      $J_5$&CuI$-\mathrm{O}_5-$CuII     & Cu-O(H)-Cu    & 3.004 & 98.53$^\circ$ & FM & FM\\
 &CuI$-\mathrm{O}_6-$CuII     & Cu-O(H)-Cu    &   & 98.75$^\circ$ & FM & black (dash)\\

      \br
   \end{tabular}}
\endfulltable

\subsection{Crystal Structure}
The crystal structure of spangolite has previously been established and extensively described by 
Hawthorne {\it et. al.}~\cite{Hawthorne:1993p1446}.  Here we discuss the crystal structure with particular reference to the magnetic properties.  Spangolite crystallizes in the trigonal space group $P3_1c$.  The structure consists of Cu$_6$Al(SO$_4$)Cl(OH)$_{12}$ layers, well separated parallel to the $c$-axis by water molecules (layers occur at $z \approx 0.0, 0.5$ with $c = 14.2995$ \AA, so are separated by 7.15 \AA).  Each layer contains a slightly distorted, $1/7$ depleted triangular lattice of metal sites occupied by Cu$^{2+}$ ions, as illustrated in Fig.~\ref{spanglatt}a.  The depletion is due to the presence of the Al$^{3+}$ ions, which form a triangular superlattice with a cell $\approx \sqrt3(3/2)$ bigger than that of the underlying triangular lattice.

We have used single crystal x-ray diffraction to confirm that our material does have this structure.  Full details of the unit cell and coordinates are given in the appendix in a form suitable for comparison with those found by Hawthorne {\it et al.}~\cite{Hawthorne:1993p1446}.  In general our coordinates agree closely ($\sim 1\%$) for the copper, sulphur, aluminum and oxygen atoms.  The main differences between the two structures are somewhat shorter lattice parameters in our case, and in the description of the interlayer water and hydrogen atom positions.  Previously a split site was found for the oxygen atom of the interlayer water molecules (O7A and O7B in their work), whereas we have a single site approximately at the average of these two positions (O7).  We have located three of the four hydrogen atoms making up the hydroxide ligands, and two for the interlayer water molecule, whereas Ref.~\cite{Hawthorne:1993p1446} has four complete hydroxide groups but no hydrogen atoms located on the interlayer water molecules.  Our hydrogen atom positions were fixed using a riding model, in which the hydrogen atoms are maintained at idealized positions relative to a parent anion.  Similarly, in Ref.~\cite{Hawthorne:1993p1446}, the hydrogen atoms had to be fixed in chemically reasonable positions to satisfy hydrogen bonding requirements.  Also in Ref.~\cite{Hawthorne:1993p1446} evidence of positional disordering of the interlayer water and sulphate groups is discussed, which further increased the problem of hydrogen atom location by x-ray diffraction.  This level of disorder is much reduced in our study.  In this context it is important to note that the original structure determination was performed using data collected at room temperature~\cite{Hawthornecom}, whereas our data were obtained at 120 K.  The consequent reduction in thermal disorder is thought to explain why a split site is not required to describe the position of the oxygen atom (although the thermal parameter remains larger than those of the other atoms) and why hydrogen atoms could be located on the interlayer water molecules.  Even with this advantage, the position of the final hydrogen atom could not be stabilized in the refinement.  Despite this, our final $R$ factor is very reasonable (4.45 \%) and there can be no doubt that the metal atom positions in our sample are as reported for spangolite.   

The slight distortion of the triangular lattice means that the Cu$^{2+}$ ions occupy two sites (both are $6c$ sites, which we label CuI and CuII), forming two sets of pure trimers.  In addition there are two other types of trimer formed by the linking of the pure trimers.  The connectivity of the copper atoms is shown in Fig.~\ref{spanglatt}, where it can be seen that there are four types of copper trimer in total, with three, two, one or zero CuI (or CuII) members.  At the centre of both the CuI and CuII trimers there is a trigonal axis, which ensures that the pure trimers are perfectly equilateral.  As shown in Table~\ref{table1}, they have closely similar Cu-Cu distances.  The slight distortion of the layer means that the two families of trimers are displaced vertically with respect to each other by 0.175 \AA.    Consequently the linking trimers are not ideal and each contains three distinct types of Cu-Cu edge.  

Fig.~\ref{spanglatt} shows that with uniform antiferromagnetic exchange interactions, the system must be highly frustrated, and it would be a realization of the maple leaf lattice discussed in Ref.~\cite{Schmalfuss:2002p3338}.  However, the above description of the lattice implies that there will be several exchange constants.  The Goodenough-Kanamori rules~\cite{goodenough} provide a framework for estimating signs and relative magnitudes of exchange constants.  Superexchange interactions between cations, mediated by an anion, will be antiferromagnetic if the subtended angle is 180$^\circ$ and ferromagnetic if it is 90$^\circ$.  Generally the antiferromagnetic exchange will be stronger, but decreases to zero and becomes ferromagnetic at some critical angle $\alpha_c$.  We now point out the salient features of the near neighbour exchange paths.

To apply the Goodenough-Kanamori rule to spangolite, we first need to estimate the crossover angle $\alpha_c$.  This depends on the coordination of the anions.  The crystallographic structure of part of a  Cu$_6$Al(SO$_4$)Cl(OH)$_{12}$ layer is illustrated in Fig.~\ref{spanglatt}b.  It can be seen that both copper and aluminum atoms lie at the center of a distorted octahedron of anions and that these octahedra share edges.  This means that each Cu-Cu pair has two bridging anions which may be OH$^-$, O(SO$_3$)$^{2-}$ or Cl$^-$, as tabulated in Table~\ref{table1}.  The hydroxide anions bridge two copper atoms and an aluminum atom, with the hydrogen projecting out of the plane of the layer.  The O(SO$_3$)$^{2-}$ or Cl$^-$ lie on the trigonal axes and bridge three copper atoms in the pure trimers, CuI in the case of chloride and CuII in the case of the sulphate group.  The geometry of the bridging atoms is known as $\mu_3$ and they can be regarded as $sp^3$ hybridized~\cite{Wills:2008p1262}.  However, the angles around the oxygen atoms indicate that they are quite strongly distorted from the ideal tetrahedral angles of $sp^3$ hybridization.  Studies of cubanes  containing [Cu$_4$($\mu_3-$OH)$_4$] involving structure-property correlation and DFT calculation have found $\alpha_c$ to be in the range 101-105$^\circ$~\cite{Gutierrez:2002p1394,Gutierrez:2000p1304,Sarkar:2007p1404}.  There are five possible exchange interactions which we name $J_i$ ($i=1,5$).  The bridging groups or superexchange paths are tabulated in Table~\ref{table1} and the corresponding interactions shown in Fig.~\ref{spanglatt}.  Comparison of the bond angles shown in Table~\ref{table1} with this value suggests that most Cu-Cu pairs will have both an antiferromagnetic and a ferromagnetic bridge and that consequently the overall exchange will be weakly antiferromagnetic.  We take the sulphate and chloride ligands to have a similar angular dependence, and in any case both these bridges have angles well below any quoted $\alpha_c$ for superexchange interactions.  Just one Cu-Cu interaction appears likely to be purely ferromagnetic.

\subsection{Magnetic Susceptibility}

\begin{figure} 
\begin{center}
   \includegraphics[trim=140 260 210 50,clip=true,scale=0.8]{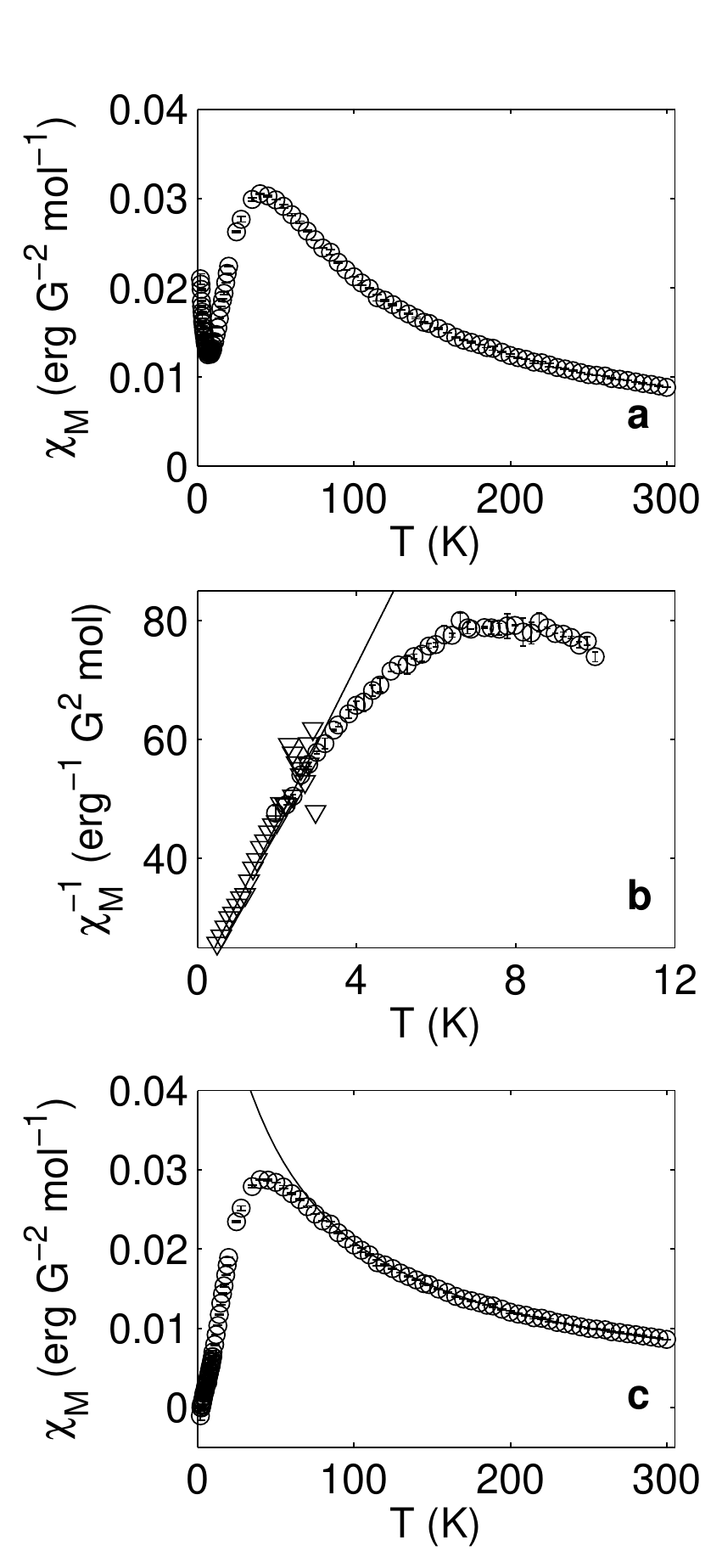} 
 \caption{The magnetic susceptibility of spangolite. a: The general features of the uncorrected susceptibility are a broad peak at $T\sim 40$ K above a minimum at $T\sim 8$ K, below this is an upturn attributed to a Curie tail.  b: The inverse susceptibility at low temperature scaled and offset to overlap with the Curie tail visible in the raw data.  The line is a Curie-Weiss law which has been fitted to both data sets across the overlapping region ($\theta_{\mathrm{CW}}=-1.434\pm0.003$ K and $C=0.0749\pm0.0001$ erg G$^{-2}$ mole$^{-1}$ K). c: The susceptibility with the extrapolated Curie-Weiss contribution from the tail subtracted.  The line is a Curie-Weiss law fitted to the data above 100 K ($\theta_{\mathrm{CW}}=-38\pm1$ K and $C=2.89\pm0.01$ erg G$^{-2}$ mole$^{-1}$ K).}
   \label{squid}
   \end{center}
\end{figure}

In Fig.~\ref{squid}a we show the general features of the magnetic susceptibility.  It is dominated by a broad maximum above a rising component at lower temperature, typical of a spin-dimer system with a Curie tail.  The low temperature data have been scaled to overlap with this tail in the 2-3 K range by subtracting a constant background, multiplying by a scaling factor and offsetting.    No satisfactory overlap is obtained if the offset is zero (i.e. the tail cannot be treated with a pure Curie law).  Instead, we used the  Curie-Weiss law $\chi=C/(T-\theta_{CW})$ to fit the inverse susceptibility of the tail to the lowest temperatures measured (0.1 K).  We extracted the Curie-Weiss temperature $\theta_{\mathrm{CW}}=-1.434\pm0.003$ K and Curie constant $C=0.0749\pm0.0001$ erg G$^{-2}$ mole$^{-1}$ K.  We obtained the effective magnetic moment from the Curie constant as $\mu_{\mathrm{eff}}=(3k_BC/N_A)^{1/2}/\mu_\mathrm{B}$ which can be compared with a value for the spin-only magnetic moment of $\mu=g(s(s+1))^{1/2}$ (since the magnetic moment is obtained from the molar susceptibility of spangolite it is in units of $\mu_{\mathrm{B}}$ per formula unit).  The moment associated with this tail is $0.772\pm0.001~\mu_{\mathrm{B}}$ f.u.$^{-1}$ implying a population of defective spins of $\approx 7.5\%$ (assuming they have $s=1/2$ and $g=2$).

\begin{figure} 
\begin{center}
  \includegraphics[trim=180 100 200 560,clip=true,scale=1.6]{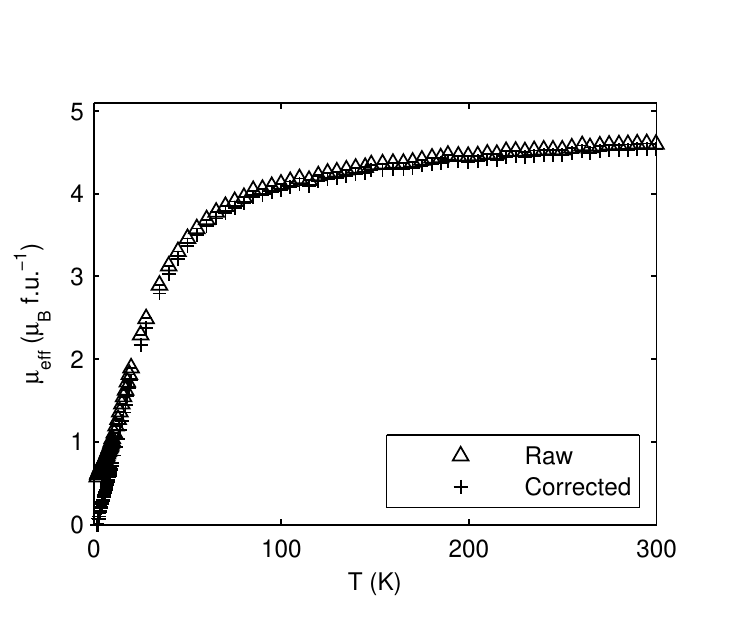} 

 \caption{The effective moment obtained from the susceptibility without the Curie tail subtracted (raw) and with this correction (corrected).  The uncorrected moment tends to a value of $\mu_{\mathrm{eff}}\sim 4.6$ $\mu_\mathrm{B}$ f.u.$^{-1}$ and the corrected value is 4.55 $\mu_\mathrm{B}$ f.u.$^{-1}$. }
   \label{mueff}
   \end{center}
\end{figure}

The extrapolated contribution from the tail can be subtracted from the susceptibility to give the intrinsic susceptibility of spangolite.  The result is shown in Fig.~\ref{squid}c, clearly exhibiting a non-magnetic (i.e. singlet) state at $T\sim 8$ K.  At high temperature the susceptibility follows the Curie-Weiss law with $\theta_{\mathrm{CW}}=-38\pm1$ K and $C=2.89\pm0.01$ erg G$^{-2}$ mole$^{-1}$ K.  The moment extracted from the Curie constant is $4.79\pm0.01~ \mu_\mathrm{B}$ f.u.$^{-1}$.  This is significantly reduced from the expected value for the spangolite formula unit, which has six $s=1/2$ and anticipated $\mu=10.39~\mu_\mathrm{B}$ f.u.$^{-1}$.  If the low temperature tail is not subtracted, a slightly larger moment of $4.85\pm 0.01$ $\mu_\mathrm{B}$ f.u.$^{-1}$ can be obtained from the Curie constant.  In Fig.~\ref{mueff} we show the effective moment obtained from the susceptibility using the expression $\mu_{\mathrm{eff}} = \sqrt{8\chi T}$.  The moment does not saturate within the temperature range of our experiment but tends to a value of 4.55 $\mu_\mathrm{B}$ f.u.$^{-1}$.  A slightly larger value of 4.6 $\mu_\mathrm{B}$ f.u.$^{-1}$ is obtained if the tail is not subtracted.

\section{Discussion and Conclusion}

The form of the susceptibility is typical of a system with a singlet groundstate.  In view of this, the most basic hypothesis of the low temperature magnetic behaviour in spangolite would be the formation of non-interacting dimers.  In this case, the high temperature moment would correspond to six $s=1/2$.  Because of the reduction in moment, the susceptibility is not reproduced by models of one or more non-interacting dimer species (i.e. the Bleaney-Bowers equation), or derivative models where the dimers interact with a mean field~\cite{singh,QuinteroCastro:2010p1983,Haraldsen:2005p1372}.  The observed and expected moments (even including the contribution from the tail) could imply that approximately half of the copper atoms had been replaced by diamagnetic substituents.   If the moment reduction is due to such a drastic level of random dilution it would seem to make the observation of such a clear transition into a singlet state unlikely.  The dilution would have to be random to preserve the crystal symmetry and so although many singlet pairs might still be formed, a much greater proportion of defective spins and consequent stronger tail would be expected.  The percolation threshold of the maple leaf lattice is 0.579498(3)~\cite{suding} suggesting that the large scale formation of a single ground state would be disrupted by such strong dilution. Furthermore, it should be visible in the crystal structure determination as a large thermal parameter associated with disorder on the copper sites, but this is not observed.

Assuming therefore that the moment reduction is an intrinsic effect, simple explanations could involve the formation of strongly  bound clusters.  For example, if the intra-trimer interactions are strong at the temperatures studied, one would observe two effective $s=1/2$, with developing interactions between trimers being responsible for the reduction in susceptibility at lower temperature.  Higher temperature measurements would then reveal a crossover to another Curie-Weiss law characteristic of the six $s=1/2$ of the formula unit.  We have not been able to verify the second part of this situation as the temperature range available to us is limited.  However, the moment we observe is, in fact, too large to agree with the first part (for two effective $s = 1/2$ per formula unit, we expect $\mu= 3.46~\mu_\mathrm{B}$ but observe $\mu= 4.79~\mu_\mathrm{B}$).  

Because the susceptibility falls to zero below a broad maximum, we conclude that the groundstate of spangolite is a type of singlet state.   It is not described by simple models of spin dimers because the high temperature moment is strongly reduced.  This moment reduction does not appear to be due to diamagnetic dilution as the level of dilution required would be expected to produce a much larger population of defective spins and greater level of disorder in the crystal structure than is observed.  While the sample is small and of natural origin, every effort has been made to control the quality of the material used and this strong moment reduction therefore appears to be an intrinsic effect, perhaps due to the formation of strongly bound trimers with an additional orbital moment.    We note that spangolite approximates to the $s=1/2$ model on the maple leaf lattice, a system with no other experimental realization, but that six-sublattice order is currently predicted for such a material.    Larger samples of high purity are a prerequisite before further conclusions can be drawn.  Finally, during the preparation of this manuscript we also became aware of Sabelliite, ideally (Cu,Zn)$_2$Zn[(As,Sb)O$_4$](OH)$_3$, which in principle would offer a single sublattice realization of the maple leaf lattice~\cite{Olmi:1995p3962}.

\ack
TF thanks Andrew Wills (UCL) for discussion, we acknowledge the EPSRC, UK for funding of work done in London (EPSRC grant EP/C534654) and Southampton (EPSRC UK National Crystallography Service), and SNF for work done in Lausanne.

\appendix
\section{Crystal structure details}

In Tables~\ref{table2} and ~\ref{table3} we collect the details of our version of the crystal structure in a form which can be compared with similar information appearing in Ref.~\cite{Hawthorne:1993p1446}.

\Table{ \label{table2}Crystallographic data for spangolite. }   
\begin{tabular}{c c c c} 
\br
$a$(\AA)         & 8.2524(3)   & Radiation                         & Mo $K_\alpha$\\
$c$(\AA)          & 14.2995(9) & Total $|I|$                        & 6115\\
$V$(\AA$^3$) & 843.36(7)   & Unique $|F_o|$                & 1308\\
Space group   & $P3_1c$    & No. of $|F_o| > 2\sigma$ & 1174\\
                       &                    & Final $R (\%)$                  & 4.45 \\ 
\br
\end{tabular}
\endTable

\Table{ \label{table3}Crystal structure data for spangolite (atom naming scheme as in Ref.~\cite{Hawthorne:1993p1446}, except HWA and HWB (interlayer water hydrogen atoms) which are not located in Ref.~\cite{Hawthorne:1993p1446}, and H3 which was not located in this work but exists in Ref.~\cite{Hawthorne:1993p1446}).  Hydrogen atom positions and thermal parameters without errors were fixed in the refinement.  The hydrogen atom positions ride on that of a parent atom (e.g. H4 is associated with O4) and can be regarded as carrying the associated error.  }   
\begin{tabular}{c  c c c c } 
\br
Atom  & $x$               & $y$               & $z$              & $U_{\mathrm{eq}}$ \\
\mr
CuI    & 0.090800(9) &0.464760(8)  & 0.000000(4) & 0.00993(2)\\ 
CuII   & 0.798400(8) & 0.040810(8) & -0.012220(5) & 0.00940(2)\\  
S         & 0                   & 0                  & -0.22354(19) & 0.02062(7)\\ 
Cl        & 1/3        & 2/3          &0.15456(2)     & 0.02658(8)\\
Al        & 2/3         & 1/3          & -0.01386(2)   & 0.00741(6) \\ 
O1     & 0.17694(7)   & 0.01926(7)    &-0.25767(3)     &0.03371(17)\\
O2      & 0.0000          & 0.0000          & -0.11936(5)    &0.01222(17)\\
O3      & 0.81044(5)   & 0.25253(5)    & -0.08320(3)    &0.01033(12)\\
O4      & 0.52523(5)   & 0.41064(5)    &0.059030(3)    & 0.01091(12)\\
O5      & 0.29725(5)   & 0.45058(5)    & -0.061070(3)   &0.01052(12)\\ 
O6      & 0.03661(5)   & 0.21901(5)    &0.04956(3)    & 0.00840(11)\\
O7B   & 0.59830(9)   & 0.08950(16)  &0.25282(5)     & 0.1113(4)\\
H4     & 0.5343          &  0.4089         & 0.12871          &0.013  \\
H5      & 0.2988          & 0.4599           & 0.86921         & 0.013  \\
H6      & 0.0351          & 0.2103          & 0.1931         &      0.010 \\
HWA & 0.8108(17)   & 0.5133(19)    & 0.25500(13)   &0.167   \\
HWB & 0.997(2)       & 0.6134(18)    & 0.2342(11)  & 0.167     \\ 

\br
\end{tabular}
\endTable

\section*{References}

\end{document}